\documentclass[final,authoryear,5p,times,12pt,twocolumn]{elsarticle}

\usepackage{amssymb}
\usepackage{amsmath}
\journal{Geothermics}

\begin{document}

\begin{frontmatter}

%% Title, authors and addresses

%% use the tnoteref command within \title for footnotes;
%% use the tnotetext command for the associated footnote;
%% use the fnref command within \author or \address for footnotes;
%% use the fntext command for the associated footnote;
%% use the corref command within \author for corresponding author footnotes;
%% use the cortext command for the associated footnote;
%% use the ead command for the email address,
%% and the form \ead[url] for the home page:
%%
%\title{Nasse Noff\tnoteref{label1}}
%\tnotetext[label1]{}
%\author{Kjetil Hals\corref{cor1}\fnref{label2}}
%\ead{kjetil.hals@cmr.no}
%% \ead[url]{home page}
 %\fntext[label2]{}
%\cortext[cor1]{}
%\address{Christian Michelsen Research, P.O. Box 6031,  NO-5892 Bergen, Norway.%$\fnref{label3}}
%% \fntext[label3]{}

\author[label1]{Kjetil M. D. Hals }
\ead{kjetil.hals@cmr.no}
\address[label1]{Christian Michelsen Research, P.O. Box 6031,  NO-5892 Bergen, Norway. }

\author[label2]{Inga Berre }
\ead{inga.berre@math.uib.no}
\address[label2]{Department of Mathematics, University of Bergen, P.O. Box 7800, NO-5020 Bergen, Norway. }

\title{Thermal Fracturing of Geothermal Wells and the Effects of Borehole Orientation}

%% use optional labels to link authors explicitly to addresses:
%% \author[label1,label2]{<author name>}
%% \address[label1]{<address>}
%% \address[label2]{<address>}

\begin{abstract}
An enhanced geothermal system (EGS) expands the potential of geothermal energy by enabling the exploitation of regions that lack conventional hydrothermal resources. The EGS subsurface system is created by engineering enhanced flow paths between injection and production wells. Hydraulic stimulation of existing fracture networks has been successfully achieved for unconventional geothermal resources. More recently proposed concepts increase the use of drilled wellbores in hard rock to connect the injection and production wells. The present work investigates the long-term thermal effects of deviated geothermal wellbores and studies how the cooling of the borehole wall results in thermally induced tensile fractures. The results show that induced fractures are created by a combination of in situ and thermal stresses, and that the extent to which thermally induced tensile wall fractures are created largely depends on how the wellbores are oriented with respect to the pre-existing stresses of the reservoir. If the system is not optimized with respect to in situ stresses, the risk of wellbore instability becomes severe within less than a year of production. In contrast, if the orientation of the wellbores is optimized, thermally induced instabilities can be completely excluded as potential risks for the operational lifetime of the system. Furthermore, our results show that the thermal failure process strongly depends on the temperature of the injected water but is only weakly affected by the injection rate.
\end{abstract}

\begin{keyword}
%% keywords here, in the form: keyword \sep keyword
Enhanced geothermal system\sep thermal fracturing\sep numerical modeling
%% MSC codes here, in the form: \MSC code \sep code
%% or \MSC[2008] code \sep code (2000 is the default)

\end{keyword}

\end{frontmatter}

% \linenumbers

%% main text
\section{Introduction}
\begin{figure}[ht] 
\centering 
\includegraphics[scale=1.0]{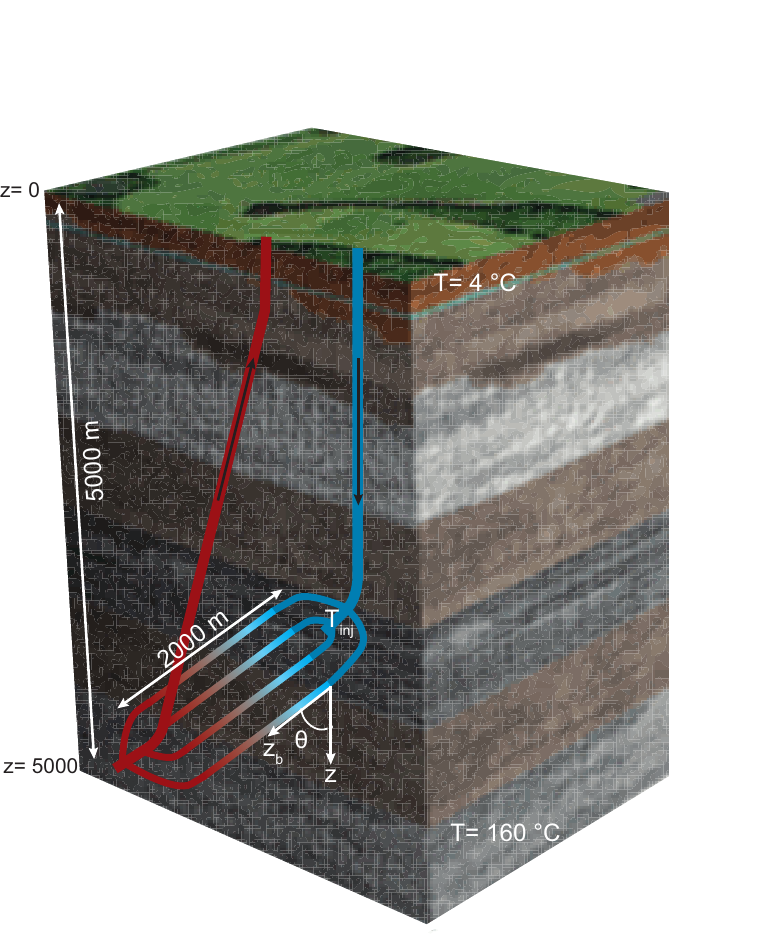} 
\caption{Illustration of a WEGS concept based on connected wellbores. The injection and production wells are interconnected by drilled wellbores (heat exchangers) that are tilted by an angle of $\theta$ with respect to the z axis. The length of each heat exchanger is $2000$ m. $z_b$ represents the z-direction in the borehole coordinate system, while $z$ is the vertical direction in the geographic coordinate system.
The maximum depth of the wells is $5000$ m below the surface. The temperature at 5000 m is assumed to be $160$ $^{\circ}$C, while the water enters the heat exchangers with a temperature of $T_{\rm inj}$. }
\label{Fig1} 
\end{figure}

Enhanced geothermal systems (EGSs) have emerged as a promising new type of geothermal power technology. An EGS uses hydraulic fracturing to create a fracture network in low-permeability rock layers to enhance the connectivity between typically deviated  injection and production wells~\citep{abe}. Still in the development stage, EGS technology has been criticized because of its potential to cause the initiation of larger seismic events~\citep{majer}. Recently, two alternative EGS concepts have been presented~\citep{holmberg,zhang} that increase the use of deviated wellbores to ensure connection between the production and injection wells (Fig.~\ref{Fig1}). One approach avoids the use of hydraulic stimulation,  thereby reducing the risk of induced seismicity, by interconnecting the production and injection wells via drilled wellbores~\citep{holmberg}. The other approach uses drilled boreholes to improve the connection between the wellbores and the existing or stimulated fracture networks~\citep{zhang}.  The present study refers to these new concepts as wellbore-based EGSs (WEGSs). The operational lifetimes of both EGSs and WEGSs largely depend on long-term effects, such as mineral scaling, corrosion, and fracturing resulting from thermal stress,
which could alter system permeability. The present article focuses on how long-term thermal effects can result in borehole instabilities due to fracturing.  

Stress is induced in a reservoir that is exposed to compression or tension. The sources of internal stress can be forces that are created due to a change of the reservoirÕs temperature or, in the case of a porous medium, a change in fluid pressure. Large pre-existing (in situ) stresses are also typically present in the reservoir due to tectonic forces and the overburden pressure~\citep{zoback_book}. Tensile or compressive rock failure is observed if the stress exceeds the rock strength~\citep{zoback_book}. Modeling this fracturing is complicated, because the process depends on several coupled mechanisms that act on different length and time scales~\citep{ahansen,hals}. Fracturing has been studied extensively over the last few decades with regard to the stabilities of deviated boreholes~\citep{perkins,peska}. In the petroleum industry, one of the major operational concerns related to drilled wellbores has been the prevention of borehole wall failure~\citep{cooper}. This type of failure is initiated by pressure and temperature gradients that are created around wellbores during drilling~\citep{perkins}. Several previously published studies have investigated the stability of inclined boreholes. In particular, pressure-induced failures have been the focus of oil-well studies. 
In addition to numerical and experimental studies~\citep{daneshy,yew,baumgartner,rawlings}, 
the problem of borehole wall failure has been studied analytically via linear elastic models~\citep{bradley,aadnoy}. \citet{peska} presented a thorough investigation of borehole stabilities based on arbitrary borehole orientations in a wide variety of in situ stress states.  They employed analytic expressions that describe the principal effective stresses at the wellbore wall of a cylindrical well while considering fluid pressure variations~\citep{hiramatsu,fairhurst}. 
Similar analytic expressions have been used in studies of thermally induced failures~\citep{myklestad, perkins, zoback_book}. 
The effects of thermal fracturing have also been of concern in carbon capture and storage (CCS) technology. Recent studies have indicated that thermally induced stresses significantly influence the safety of the seal and the fracturing of the caprock~\citep{luo,goodarzi_2010,goodarzi_2012}. 
In the context of geothermal reservoirs, several studies have focused on thermally induced stresses, mainly concentrating on heat extraction enhancements due to thermal cracks~\citep{murphy,demuth}. Thus far, 
few works have attempted thorough stability analyses of geothermal wells that investigate the interplay between in situ stresses, borehole orientations, and long-term thermal effects.  

The present paper combines numerical and analytical methods in the
first theoretical study of the long-term thermal effects of deviated geothermal wellbores in regions where convective fluxes between geothermal reservoirs and wellbores are negligible. Temperature evolution in the geothermal well and rock reservoir are simulated using a numerical COMSOL Multiphysics model~\citep{comsol}, while stresses at the wellbore wall are calculated from analytic expressions valid for cylindrical wells. We show that the degree of thermal tensile fracturing is extremely sensitive to how the drilled wellbores are oriented with respect to in situ stress states. In the case of a large horizontal stress, the wellbores should be oriented along the maximum horizontal stress direction. This orientation nearly eliminates the risks for thermally induced borehole instabilities, even after several years of injecting cold water. In contrast, when the wellbores are oriented along the minimum stress direction, wellbore instabilities are observed within less than one year. 
In addition, we show that the degree of thermal fracturing strongly depends on the temperature of the injected water but is only weakly affected by the injection rate.
The strong fluid-temperature dependency of the thermal failure process allows one to control the degree of fracturing for a given in situ stress state.
Our results provide important insights for optimizing WEGSs as well as oil and geothermal wells, in general.  We expect that the results can be applied to CCS technology to enhance the safety of geological storage sites.  

This paper is organized as follows. In Section~\ref{Sec:theory}, we present the theory and governing equations that model the geothermal system. Section~\ref{Sec:model} provides a description of the model system under consideration, while Section~\ref{Sec:results} presents our findings. We conclude and summarize our results in Section~\ref{Sec:summary}.

\section{Governing Equations}\label{Sec:theory}
The following sections, present the equations that govern the temperature evolution of the reservoir and the heat transfer process between the hot rock and cold water that circulates through the wellbores. In addition,  the analytic expressions for the stresses that act on the wall of deviated, cylindrical wellbores in the presence of in situ stresses, fluid pressure, and thermal stresses are presented.     

\subsection{Temperature Equations}
The temperature equations for the system considered herein are derived from the conservation of energy, which is described by two coupled processes: the energy change of the fluid and the energy change of the rock. These two processes are coupled by the heat transfer that occurs at the borehole wall. This heat transfer is a complicated process that depends on several different parameters, including the velocity and the temperature of the fluid, the mineral composition of the rock, and the wellbore properties. The system is
characterized by the following two temperature equations that describe the temperature, $T_f(\mathbf{r},t)$,  of the fluid inside the wellbore, $\Omega_w$, and the temperature, $T_r(\mathbf{r},t)$, of the rock in the region, $\Omega_r$, outside the wellbore ~\citep{bejan}:
\begin{equation}
\rho_f c_{p,f}\left(   \frac{\partial T_f}{\partial t}  +  \mathbf{v}\cdot\boldsymbol{\nabla} T_f \right) =  \boldsymbol{\nabla}\cdot\left( \kappa_f \boldsymbol{\nabla} T_f \right), \  \mathbf{x}\in \Omega_w, \label{Eq:T_f} 
\end{equation}
\begin{equation}
\rho_r c_{p,r} \frac{\partial T_r}{\partial t} =  \boldsymbol{\nabla}\cdot\left( \kappa_r \boldsymbol{\nabla} T_r \right) ,  \ \mathbf{x}\in \Omega_r   \label{Eq:T_r} .
\end{equation}
Here, $\rho_f$ ($\rho_r$), $c_{p,f}$ ($c_{p,r}$), and $\kappa_f$ ($\kappa_r$) are the density, isobaric specific heat capacity, and thermal conductivity of the fluid (rock), respectively,  while $\mathbf{v}$ is the fluid velocity. 
These two equations are coupled by the heat transfer between the fluid and rock. To lowest order in the temperature difference $(T_f - T_r)$, the heat flux across the borehole wall is given by~\citep{bejan}:
\begin{equation}
\mathbf{q} = h (T_{\infty,f} - T_s)\mathbf{n}, \label{Eq:HeatFlux}
\end{equation}
where $\mathbf{n}$ is the surface normal of the wall pointing into the rock reservoir, $h$ is the heat transfer coefficient, $T_s$ is the temperature of the borehole wall, and $T_{\infty,f}$ is the bulk temperature of the fluid in the wellbore. 
The heat transfer coefficient can be expressed in terms of the Nusselt number, $Nu$, of the fluid by the following relation~\citep{bejan}:   
\begin{equation}
Nu  = \frac{h D_h}{\kappa_f}.
\end{equation}
Here, $D_h$ is the hydraulic diameter, which is equal to the borehole diameter, $D$, for a cylindrical wellbore. 
Several empirical models for the Nusselt number exist. In the present study, we calculate $Nu$ from the Dittus-Boelter equation (see Section~\ref{Sec:model} for further details).  
   
\subsection{Effective Stresses and Failure Criterion} 
In real rock reservoirs, in situ stresses are present. The instability of a borehole largely depends on its orientation with respect to in situ principal stress directions. In the following analysis, we present analytic expressions for the effective stresses on the borehole wall for arbitrary borehole orientations and principal in situ stress directions. Further details can be found in \citet{peska} and \citet{zoback_book}, and the references therein.

Let $S_{ij}$ denote the in situ stress tensor represented in the borehole coordinate system, where the $z_b$ axis points along the borehole axis (Fig.~\ref{Fig1}), the $x_b$ axis is in the plane perpendicular to the borehole axis (oriented towards to the bottom side of the borehole), and the $y_b$ axis is in the same plane, but is perpendicular to $x_b$ (see \citet{peska} for an illustration). The effective stress tensor in the borehole coordinate system is then, $\sigma_{ij}= S_{ij} - bP_p\delta_{ij}$, where $P_p$ is the pore pressure, $b$ is the Biot-Willis parameter, and $\delta_{ij}$ is the Kronecker delta. A positive stress implies that the system is under compression. The effective stresses at the borehole wall of a cylindrical wellbore are:
\begin{eqnarray}
\sigma_{z_b z_b} &=& \sigma_{33} - 2\nu \left(  \sigma_{11} - \sigma_{22} \right) \cos ( 2\tilde{\theta})  -  \nonumber \\
& & 4\nu\sigma_{12}\sin (2 \tilde{\theta}) , \nonumber 
\end{eqnarray}
\begin{eqnarray}
\sigma_{\tilde{\theta}\tilde{\theta}} &=& \sigma_{11} + \sigma_{22} - 2 \left(  \sigma_{11} - \sigma_{22} \right) \cos ( 2\tilde{\theta} ) - \nonumber \\ 
& & 4\sigma_{12}\sin ( 2 \tilde{\theta} ) - \Delta p    - \sigma_{\tilde{\theta}\tilde{\theta}}^{\Delta T}  , \nonumber 
\end{eqnarray}
\begin{equation}
\tau_{\tilde{\theta}z_b} =   2\left( \sigma_{23} \cos ( \tilde{\theta})  - \sigma_{13} \sin ( \tilde{\theta}) \right)   , \nonumber 
\end{equation}
\begin{equation}
\sigma_{rr} =   \Delta p  + \sigma_{rr}^{\Delta T} . \label{Eq:EffStress}
\end{equation}
Here, $\tilde{\theta}$ is the angle (around the borehole) measured counterclockwise from the $x_b$ axis, $r$ is the radial direction away from the center of the cylindrical well, 
 $\Delta p$ is the difference between the borehole fluid pressure and pore pressure in the rock, and $\sigma_{rr}^{\Delta T}$ and $\sigma_{\tilde{\theta}\tilde{\theta}}^{\Delta T} $
 are the thermally induced stresses. In the present study, we are only concerned with the value of  $\sigma_{\tilde{\theta}\tilde{\theta}}^{\Delta T} $, which can be reduced to the following expression in the steady state~\citep{zoback_book}:
 \begin{equation}
\sigma_{\tilde{\theta}\tilde{\theta}}^{\Delta T} =    \alpha_ t E \Delta T / (1 - \nu).
\end{equation}
Here, $\nu$ is the Poisson's ratio, $\alpha_t$ is the linear coefficient of thermal expansion, $E$ is the Young's modulus, and $\Delta T (t)\equiv T_{bw, i} - T_{bw}(t)$ is the difference between the initial temperature, $T_{bw, i}$, of the borehole wall and the time-dependent temperature, $T_{bw}(t)$, of the heated or cooled borehole wall. Although the above equations assume that $\Delta p$ and $\Delta T$ are constant along the borehole axis, the expressions constitute good approximations for more general situations if the spatial variations of $\Delta T$ and $\Delta p$ along the borehole axis are weak compared to the spatial variations along the radial direction. 

From Eq.~\eqref{Eq:EffStress}, we see that increasing the borehole fluid pressure or cooling the rock reduces the tangential stress $\sigma_{\tilde{\theta}\tilde{\theta}}$. For sufficiently large values of  $\Delta p$ and $\Delta T$, the system may evolve from a compressive state into a tensile state. Because the tensile strength of rock  is relatively weak, a common tensile failure criterion (for a fixed $\tilde{\theta}$ direction) is when the minimal effective stress in the plane tangential to the borehole becomes negative~\citep{zoback_book}.   
The minimal effective stress is given by:
\begin{equation}
\sigma_{t~min} =  \frac{1}{2} \left(  \sigma_{z_b z_b} + \sigma_{\tilde{\theta}\tilde{\theta}}  - \sqrt{ ( \sigma_{z_b z_b} - \sigma_{\tilde{\theta}\tilde{\theta}})^2 + 4 \tau_{\tilde{\theta}z_b} ^2}    \right) . \label{Eq:MinStress}
\end{equation}
Therefore, the present analysis assumes the creation of tensile wall fractures when $\sigma_{t~min} < 0 $.

\section{Model Description}\label{Sec:model}
This section provides a description of the model system used herein to study of thermally induced instabilities in geothermal wells.

\subsection{COMSOL Multiphysics Model}
A COMSOL Multiphysics (version 4.3) model is used
to simulate the temperature evolution of a deviated wellbore cooled by injected fluid. In the case of multiple wellbores, we assume that the separations between the boreholes are sufficiently large to avoid thermal interaction, such that each borehole can be modeled separately. We only simulate the temperature evolution along a 2000m-long wellbore segment. In the WEGS concept illustrated in Fig.~\ref{Fig1}, such a wellbore segment is provided by one of the inclined boreholes that connects the injection and production wells. The COMSOL model of the 2000m-long borehole segment is illustrated in Fig~\ref{Fig2}.
The system is modeled in the borehole coordinate system, in which 
the location of the well segment is always between $3000~m \leq z_b \leq 5000~m$, irrespective of the tilting angle, $\theta$.
The system is assumed to be rotational symmetric around the $z_b$ axis, and the radius is 50 m.  
Rotational symmetry is also a good approximation in the case of a tilted wellbore, $\theta > 0$, because the difference between the temperature at the center of the system and at the outer surface (for a fixed $z_b$ value) is $\sim 50\sin(\theta) dT/dz$. For the temperature gradient, $dT/dz$, considered here, this situation yields a negligible temperature difference (less than $1.5$~K). 
The diameter of the wellbore is 0.1 m~\citep{holmberg,zhang}. 
The outlet of the wellbore segment is always located $z= 5000$ m below the surface (in the geographic coordinate system), while the depth of the inlet depends on the orientation of the borehole and is related to the tilting angle by: $z= 3000 + 2000(1- \cos\theta) $ m. 
The temperature evolution of the fluid  and rock matrix are described by Eq.~\eqref{Eq:T_f} and Eq.~\eqref{Eq:T_r}, respectively. 
The injected fluid is  water.  The fluid properties, including the density, thermal conductivity, and heat capacity, are given by the built-in
COMSOL functions for water.  The fluid velocity, $\mathbf{v}$, is assumed to be constant throughout the entire wellbore and is calculated from the injection rate and fluid density at a temperature of 60 $^{\circ}$C. 
The rock matrix has the following properties, which are typical for Fennoscandia~\citep{slagstad}: the heat capacity is $c_{p,r}= 850$ J/kgK, the density  is $\rho_r = 2600$ kg/$\rm m^3$, and the thermal conductivity is $\kappa_r = 2.8$ W/mK.
The initial temperature, $T_i(z_b)$, of the rock reservoir in Fig.~\ref{Fig2} depends on how the wellbore is tilted with respect to the vertical $z$ axis and is given by:
\begin{figure}[ht] 
\centering 
\includegraphics[scale=1.0]{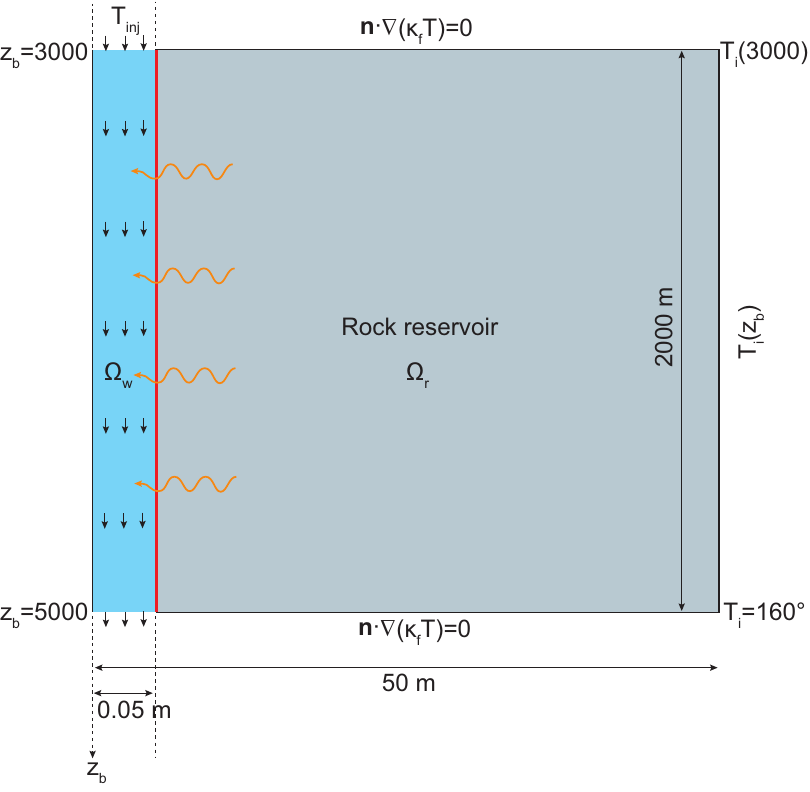} 
\caption{Illustration of the COMSOL model. The system is rotational symmetric around the $z_b$ axis. The model consists of a rock reservoir with a radius of 50 m. The drilled wellbore, which has a diameter of 0.1 m, is located at the center of the reservoir.
The borehole and rock reservoir are separated by a thin, thermally resistive layer (red line). The system is thermally insulated at the upper and lower surfaces  (i.e., $\mathbf{n}\cdot \boldsymbol{\nabla}(\kappa_f T) = 0$, where $\mathbf{n}$ is the surface normal). The outer surface temperature is assumed to be the initial temperature, $T_i (z_b)$, of the reservoir. The injected water has a temperature of $T_{\rm inj}$ and maintains a uniform velocity profile through the wellbore. }
\label{Fig2} 
\end{figure}
\begin{eqnarray}
T_i(z_b) &=& T_0 + \frac{dT}{dz_b} z_b, \\
T_0 &=& T_s + 5000(1- \cos(\theta))\frac{dT}{dz}, \nonumber \\
\frac{dT}{dz_b} &=& \frac{dT}{dz}\cos(\theta). \nonumber 
\end{eqnarray}
Here, $\theta$ is the tilting angle, $T_s= 4^{\circ}$ is the surface temperature, and $dT/dz = 0.0312$ K/m is the temperature gradient. 
The water in the wellbore is initially assumed to be in thermal equilibrium with the rock (before the simulation starts and cold water is injected into the wellbore with a temperature of $T_{\rm inj}$). 
The boundary conditions are shown in Fig.~\ref{Fig2}. The upper and lower surfaces of the rock matrix are thermally insulated; thus, it is assumed that the thermal currents along these two surfaces are oriented along the radial direction. 
To model a rock system that is in contact with an infinite rock reservoir,  the temperature along the outer surface is set equal to the initial temperature of the reservoir. 

Heat transfer between the rock and fluid is modeled by a thin, thermally resistive layer
at the borehole wall, such that the heat flux across the layer is given by: 
\begin{equation}
\mathbf{q} =  \kappa_f \frac{ T_f  - T_r}{d}   \mathbf{n} . \label{Eq:thermlayer}
\end{equation}
$T_f$ and $T_r$ are the temperatures of the fluid and rock on each side of the thermally resistive layer, respectively, and $d$ is the thickness of the layer. The thickness is determined by equating Eq.~\eqref{Eq:thermlayer} and Eq.~\eqref{Eq:HeatFlux},
which yields $d= D/Nu$.  The Dittus-Boelter equation, known to be valid for fully developed, turbulent flow, is used to calculate the Nusselt number, as follows~\citep{bejan}:
\begin{equation}
Nu = 0.023 Re^{0.8} Pr^{0.4}.
\end{equation}      
$Re$  and $Pr$  are the Reynolds number and Prandtl number of the fluid, respectively. We assume that these two numbers and the heat transfer coefficient are constant along the wellbore axis, and are determined by the fluid velocity and fluid properties evaluated at a temperature of 60 $^{\circ}$C. 

The main interest of the present study is the stability of the wellbore. As explained in \citet{zoback_book}, an unstable wellbore is defined as a borehole in which failure produces so much failed material from around the wellbore that it cannot be eliminated by mud circulation.  
To measure the instability, we calculate the cooling of the rock located one borehole diameter (i.e., 10 cm) into the rock reservoir (away from the borehole wall) and use this temperature decrease to check the failure criterion in Eq.~\eqref{Eq:MinStress}.
If the failure criterion is fulfilled, we expect a failure zone with a maximum thickness of approximately 10 cm around the wellbore. In this case, the thermal fracturing can potentially produce failed materials with sizes on the order of the borehole diameter.
Therefore, as a borehole instability criterion, we check to see whether the failure criterion in Eq.~\eqref{Eq:MinStress} is fulfilled at a depth of one borehole diameter into the rock.  

To examine the convergence of the COMSOL program, we performed a simulation with a double-resolution grid and found that the difference between the two temperature solutions was less than 1$\%$. 

\subsection{In Situ Stress State}
Consider a reservoir under the large horizontal tectonic stresses typical in the studies of Fennoscandia~\citep{stephansson}.
The maximum ($\sigma_H$) and minimum ($\sigma_h$) principal horizontal stresses are given by~\citep{fejerskov, stephansson}:
\begin{eqnarray}
\sigma_H (z) &=& 2.8 + 0.04 z \  \  {\rm MPa},  \\
\sigma_h (z) &=& 2.2 + 0.024 z \  \   {\rm MPa}.
\end{eqnarray} 
The maximum horizontal stress is oriented northwest/southeast (see the red, dotted line in Fig.~\ref{Fig3}). 
The vertical stress is given by $\sigma_h/\sigma_V = 1.1$. 

We assume that the water in the wellbore diffuses into the low-permeability rock surrounding the boreholes. 
Bernoulli's principle states that the velocity of the fluid results in a pressure decrease. The velocity yields a relative correction to the hydrostatic
pressure (on the order of $v^2 / g z$). At the depths (i.e., greater than 3000 m) and fluid velocities (on the order $1$ m/s) considered in this paper, the relative velocity correction to the fluid pressure is on the order $10^{-4}$.
We, therefore, neglect the difference between the borehole fluid pressure and pore pressure in the rock, i.e., $\Delta p = 0$,  and set the pore pressure in the effective stress tensor equal to
the hydrostatic pressure, i.e., $P_p = \rho_f g z$, where $g$ is the gravitational acceleration.     

We use values for the Young's modulus and Poisson's ratio that are typical for Fennoscandia~\citep{stephansson} and a thermal expansion coefficient that corresponds to a silica content of $73\%$~\citep{zoback_book}:  
$E= 6.0\cdot 10^4$ MPa, $\alpha_t = 8.5\cdot 10^{-6}$ $\rm K^{-1}$,  and $\nu = 0.2$. For simplicity, the Biot-Willis parameter is set to one  ($b=1.0$).

\section{Results}\label{Sec:results}
\begin{figure}[ht] 
\centering 
\includegraphics[scale=1.0]{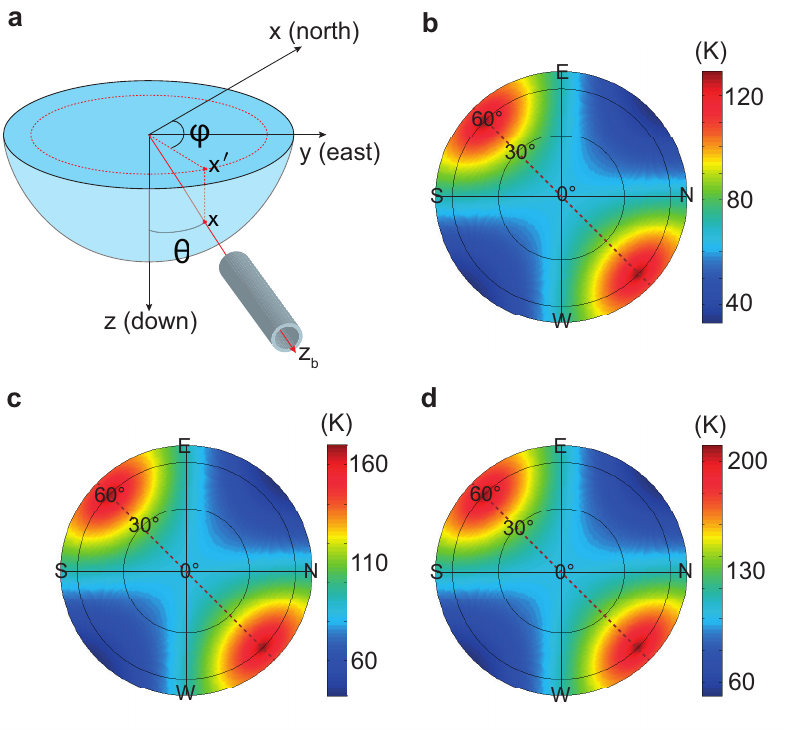} 
\caption{ ({\bf a}) All possible wellbore orientations are represented as points on the unit disk. A well that intersects the lower hemisphere of the unit sphere at the point $x$, which is parameterized by the angles $\theta$ and $\phi$, is mapped to the point $x^{'}$ by a projection onto the $xy$ plane.
({\bf b}) The temperature change of the rock (at the borehole wall) required to induce thermal fractures for all possible orientations of the wellbore at a depth of 3000 m.  
({\bf c}) The temperature change of the rock (at the borehole wall) required to induce thermal fractures for all possible orientations of the wellbore at a depth of 4000 m.  
({\bf d}) The temperature change of the rock (at the borehole wall) required to induce thermal fractures for all possible orientations of the wellbore at a depth of 5000 m.
In ({\bf b}) - ({\bf d}), the red dotted line shows the direction of the maximal horizontal stress, while the inner (outer) black circle represents wellbores that are tilted by an angle of 30$^{\circ}$ (60$^{\circ}$) with respect to the z axis.  }
\label{Fig3} 
\end{figure}

First, consider the minimum cooling of the borehole wall required to induce thermal tensile fractures. When the wellbore is oriented along one of the principal stress directions and $\Delta p_ f = \Delta T =0$, the minimum value of the effective stress in Eq.~\eqref{Eq:MinStress} becomes $\sigma_{t~min} (\tilde{\theta}_{\rm min})= 3\sigma_{22} - \sigma_{11}$. A wellbore that is directed along the $\sigma_h$, $\sigma_H$, and $\sigma_V$ directions, results in the minimum effective stresses: 
\begin{eqnarray}
\sigma_{\rm min}^h &=& 3\sigma_V - \sigma_H \\
\sigma_{\rm min}^H &=& 3\sigma_V - \sigma_h \\
\sigma_{\rm min}^V &=& 3\sigma_h - \sigma_H
\end{eqnarray}  
For the principal stress values and depths considered here, we find that $\sigma_{\rm min}^h < \sigma_{\rm min}^V < \sigma_{\rm min}^H$. Thus, we expect thermal failure to be most dominant in wellbores that are oriented along the minimum horizontal stress direction. This premise is verified by Fig.~\ref{Fig3}, which shows the minimum $\Delta T$ that induces thermal failures for all possible wellbore orientations. The results are calculated from Eq.~\eqref{Eq:MinStress} for three different depths: 3000 m, 4000 m, and 5000 m below the surface. The secant method is used to find the $\Delta T$ value that yields $\sigma_{t~min}=0$. We see that the magnitude of $\Delta T$ depends strongly on how the wellbore is oriented with respect to in situ stresses. As expected from the above arguments, the minimum $\Delta T$ appears for wellbores that are directed along the minimum principal stress direction. Wellbores that are tilted along the maximum stress direction require the largest values of $\Delta T$ to initiate thermal failure. The maximum $\Delta T$, which is above 130 $^{\circ}$C for depths greater than 3000 m, is more than three times the minimum $\Delta T$. The large $\Delta T$ values along the $\sigma_H$ direction imply that thermal failure is completely absent for wellbores oriented along this axis. In contrast, for wellbores oriented along the minimum stress direction, thermal fracturing is expected for temperature changes in the range of 40-60 $^{\circ}$C. In deep geothermal systems with reservoir temperatures greater than 140 $^{\circ}$C, a temperature change of this order is feasible. Thus, for the model system described in Section~\ref{Sec:model}, we expect thermal fracturing to be present for boreholes tilted along the minimum stress direction.

\begin{figure}[ht] 
\centering 
\includegraphics[scale=1.0]{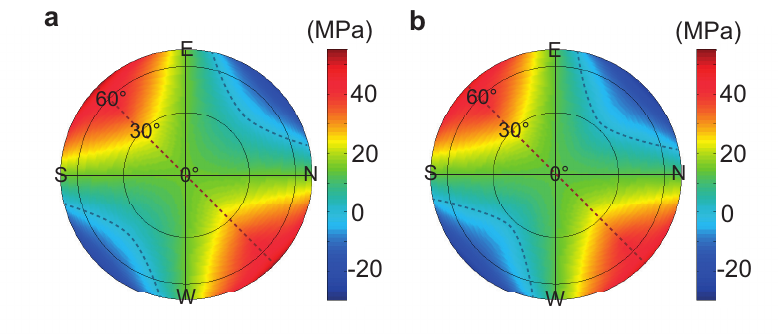} 
\caption{ ({\bf a}) The figure shows the minimum value (along the well segment) of the wellbore instability criterion for all possible orientations of the wellbore after one year of operation. 
({\bf b}) The figure shows the minimum value (along the well segment) of the wellbore instability criterion for all possible orientations of the wellbore after 20 years of operation. 
In both plots, the red dotted line shows the maximum horizontal stress direction.  In each plot, the two pockets that are confined by the blue dotted lines represent the orientations for which we expect thermally induced borehole instabilities. The injection temperature of the water is  45 $^{\circ}$C, and the injection rate is 15 L/s.   
 }
\label{Fig4} 
\end{figure}

To study thermally induced wellbore instabilities, we performed a numerical COMSOL simulation of the reservoir's temperature evolution. Fig.~\ref{Fig4} shows the minimum value of the wellbore instability criterion along the 2000m-long wellbore segment for all possible orientations of the wellbore after one year and 20 years of geothermal production. In each figure, the blue lines confine two blue pockets that represent the \emph{critical orientations} for which thermally induced instabilities  are present. Within less than one year of geothermal production, borehole instabilities occur for the wellbores that are tilted by an angle of $\theta > 45^{\circ}$ along the minimum stress direction. After 20 years of production, this critical region expands to include wellbores that are tilted by an angle of approximately  $\theta > 35^{\circ}$. Thus, for the most part, the set of critical borehole orientations is mainly fixed after one year of production, and only minor changes in the set are observed after 20 years of production because the cooling of rock close to the borehole begins as a rapid process and quickly equilibrates towards a quasi-stationary process in which the temperature changes slowly.    

\begin{figure}[ht] 
\centering 
\includegraphics[scale=1.0]{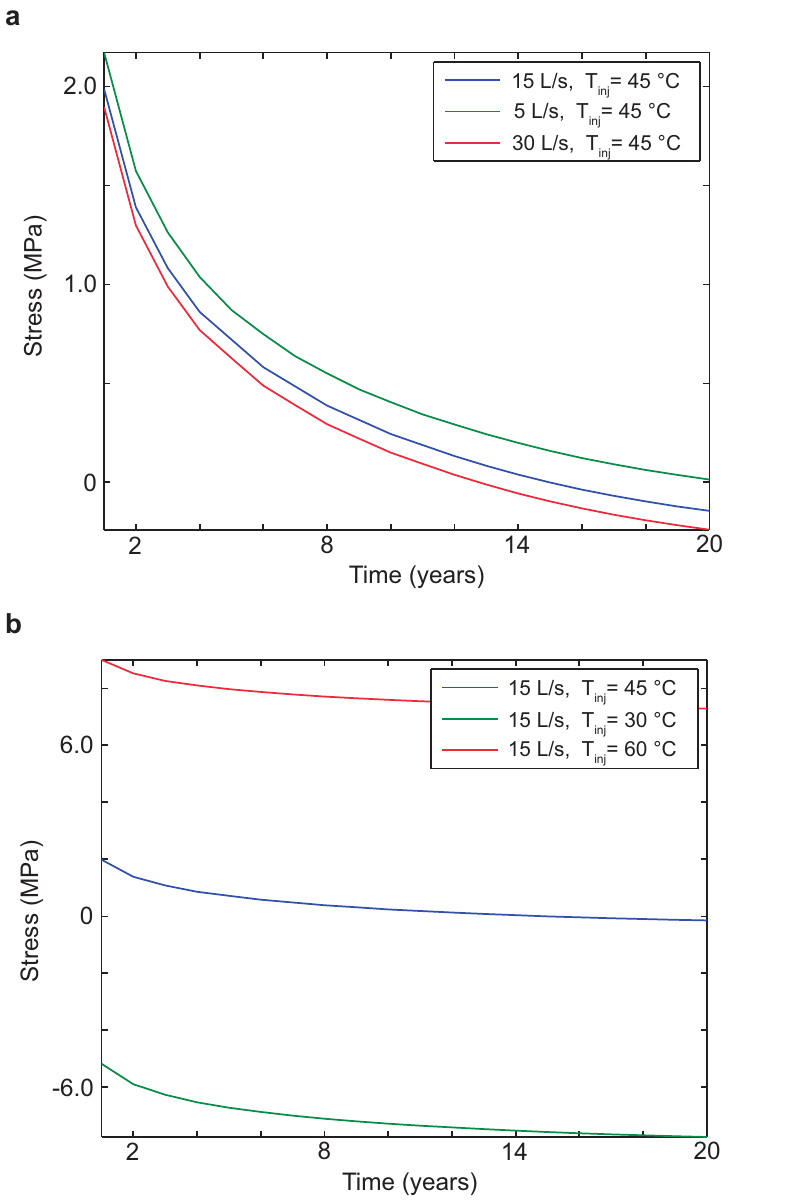} 
\caption{ ({\bf a}) The figure shows the time dependence of the minimum value of the wellbore instability criterion along the wellbore for injection rates of 5, 15, and 30 L/s. The injection temperature of the water is  45 $^{\circ}$C. 
({\bf b}) The figure shows the time dependence of the minimum value of the wellbore instability criterion along the wellbore for injection temperatures of 30, 45, and 60 $^{\circ}$C.  The injection rate is 15 L/s.
In both plots, the wellbore has an orientation of $\theta = 60^{\circ}$, $\phi = 75^{\circ}$. 
}
\label{Fig5} 
\end{figure}

We also studied of how the instability criterion depends on the injection rate and injection temperature of the fluid. Fig.~\ref{Fig5}a shows the time evolution of the minimum value of the wellbore instability criterion along the 2000m-long borehole segment for injection rates of 5, 15, and 30 L/s. The wellbore is oriented along the direction $\theta = 60^{\circ}$, $\phi= 75^{\circ}$ and the injection temperature is 45 $^{\circ}$C. The figure shows that the value of the stability criterion increases weakly with decreasing injection rate. This result is logical, given that a lower injection rate implies a lower rate of heat extraction from the reservoir. The physical reason for this weak injection-rate dependency is that the heat extraction rate is predominantly governed by the heat transfer coefficient, $h$, which scales with the Reynolds number of the fluid as $h\propto Re^{0.8}$. Fig.~\ref{Fig5}b shows the time evolution of the minimum value of the wellbore instability criterion along the 2000m-long borehole segment for injection temperatures of 30 $^{\circ}$C, 45 $^{\circ}$C, and 60 $^{\circ}$C. The orientation of the wellbore is the same as in Fig.~\ref{Fig5}a, and the injection rate is 15 L/s. We see that the value of the instability criterion depends strongly on the injection temperature of the fluid. The fluid-temperature dependency is mainly determined by the elasticity parameters of the reservoir. For the model system studied in the present paper, a temperature change of 15 K yields a thermal stress change on the order of $\alpha_t E 15 / (1 - \nu)= 9.6$ MPa.
This estimate is in good agreement with the injection-temperature dependency shown in Fig.~\ref{Fig5}b. As a consequence of this large thermal stress-change,  the set of critical borehole orientations is strongly affected by the injection temperature of the fluid. Adding or subtracting a thermal stress of $\sim 9$ MPa in Fig.~\ref{Fig4} provides an estimate for the case where the injection temperature is  $60$ $^\circ$C or $30$ $^\circ$C. Even when the injection temperature is as high as $60$ $^\circ$C, a significant number of borehole orientations become unstable.    

Our numerical simulations reveal two import considerations regarding long-term thermal effects on EGS wellbores. First, the system is extremely sensitive to how the wellbores are drilled with respect to the in situ stress state. For boreholes that are oriented along one of the critical directions, thermally induced fractures with risk of causing wellbore instabilities are observed after only a short time of production via the geothermal system. In other words, prior knowledge of the reservoir's stress state is crucial for optimal design. Second, the extent to which thermal fracturing appears is largely determined by the temperature of the injected fluid. This strong injection-temperature dependency opens a path for controlling  the thermal failure process. Our results provide important insights regarding long-term thermal effects that are essential for optimal engineering of WEGSs as well as geothermal wells and oil wells, in general.      
Importantly, our results can also be interpreted in terms of CCS technology. One of the major concerns regarding CCS technology is fracturing of the caprock caused by the CO$_2$ injection~\citep{zoback}. Recent studies have indicated that the thermal stresses  induced by injecting cold CO$_2$  play an important role in determining whether the caprock will fracture \citep{luo,goodarzi_2010,goodarzi_2012}. 
Our results imply that drilling through the caprock along an optimal direction reduces the risk of thermal caprock fracturing close to the well.   

\section{Conclusions}\label{Sec:summary}
This paper contains a theoretical study of thermal fracturing and considers the effects of borehole orientation in deviated geothermal wellbores.  We found that the risk for thermally induced wellbore instabilities is highly sensitive to how the wellbores are oriented with respect to the in situ stress state of the rock reservoir. Thermal failures can be completely eliminated if the system is optimized with respect to the principal stress directions. In contrast, if the system is not optimized, thermally induced fractures with risk of causing instabilities are observed after less than one year of geothermal production. In addition, we found that the degree of thermal fracturing depends strongly on the temperature of the injected water, while it is only weakly dependent on the injection rate of fluid. This strong fluid-temperature dependency provides a direct way to control the degree of thermal fracturing for a given reservoir stress state.

\section{Acknowledgments}
We are grateful to K. E. Brun and K. Midtt\o mme at Christian Michelsen Research for stimulating discussions regarding the COMSOL model and for Fennoscandia stress data.

\bibliographystyle{elsarticle-harv}
\bibliography{<your-bib-database>}

%% Authors are advised to submit their bibtex database files. They are
%% requested to list a bibtex style file in the manuscript if they do
%% not want to use elsarticle-harv.bst.

%% References without bibTeX database:

\end{document}